\documentclass[review]{elsarticle}

\usepackage{lineno,hyperref}
\modulolinenumbers[5]

\journal{Journal of \LaTeX\ Templates}

\usepackage{color} 
\usepackage{ulem} 
\usepackage{placeins}
\usepackage{ulem} 
\usepackage{booktabs}
\usepackage{blindtext}
\makeatletter
\newcommand{\colorcaption}[2][]{%
  \begingroup%
  \renewcommand{\@caption@fignum@sep}{ (Color online). }%
  \caption[#1]{#2}%
  \endgroup%
}
\makeatother
\newcommand\T{\rule{0pt}{3ex}}       
\newcommand\B{\rule[-1.5ex]{0pt}{0pt}} 
\usepackage{amssymb}
\usepackage{graphicx}
\usepackage{rotating}
\usepackage{tikz}
\begin{document}
\begin{frontmatter}
\title{Shell-model description for the first-forbidden $\beta ^{-}$ decay of $^{207}$Hg into the one-proton-hole nucleus $^{207}$Tl}
\author{Anil Kumar}
\ead{akumar5@ph.iitr.ac.in}
\author{ Praveen C. Srivastava}
\ead{ praveen.srivastava@ph.iitr.ac.in}
\address{Department of Physics, Indian Institute of Technology Roorkee, Roorkee 247667, India}

%
%
%
%
%
%
%
\begin{abstract}
In this work, we have performed large-scale shell-model calculations for
the first-forbidden $\beta ^{-}$ decay of $^{207}$Hg into the one-proton-hole
nucleus $^{207}$Tl corresponding to the recently available experimental
data from ISOLDE-CERN [T. A. Berry \textit{et al.,} Phys. Rev.
C 101, 054311 (2020)]. We have used the one-particle one-hole ($1p$-$1h$)
truncation for both protons and neutrons simultaneously across the doubly-shell
closure at $^{208}$Pb in the final states of $^{207}$Tl. In our calculations,
we have also considered the effect of mesonic enhancement
$\epsilon _{\mathrm{mec}}=2.01\pm 0.05$ in the rank-0 for the axial-charge matrix
element $\gamma _{5}$. Here, we have calculated the $\log ft$ values from
the ground-state of $^{207}$Hg to the several excited states of
$^{207}$Tl and obtained a good agreement between the calculated and the
experimental data. In the experimental data spin and parity for some states
are not yet confirmed, thus based on the shell-model results for the
$\log ft$ values we have given the prediction for these states. This is
the first theoretical calculation for the $\log ft$ values for these transitions.
\end{abstract}
%
\begin{keyword}
Shell-model, First-forbidden beta decay
\end{keyword}
\end{frontmatter}
\section{Introduction}
\label{sec1}

In the recent past, several experiments have been performed to study the
structure including collectivity and $\beta $ decay properties in the lead
region. In this region, the low-lying states are characterized by the single-particle
structures, while the higher energy states are due to core breaking. The
first excited $3^{-}$ state in $^{208}$Pb is highly mixed and has complex
wave function \cite{Brown2000,Podolyak2015}. Wilson et al.
\cite{Wilson2015} have measured the high-spin states above the 3.813 MeV
in the nucleus $^{207}$Tl and also they have reported the shell-model results
corresponding to $t=1$ truncation (where $t$ is the number of nucleons
excitation across the $^{208}$Pb core) for both protons and neutrons. After
that, the same group measured the new excited states in $^{207}$Tl through
the $\beta ^{-}$ decay of $^{207}$Hg in the energy range of 2.6-4.0 MeV
\cite{Berry2020}. In these experiments, several new excited states have
been identified in $^{207}$Tl. The high-precision $\beta $ spectrum of
the first-forbidden nonunique $\beta ^{-}$ decay of $^{210}$Bi has been
measured in Ref.~\cite{Alekseev2020}. The experimental evidence for competition
between allowed and first-forbidden $\beta $ decay in the case of
$^{208}$Hg${\rightarrow }^{208}$Tl to understand the mechanism in the rapid
neutron capture process is reported in Ref.~\cite{Carroll2020}. The aims
of these experiments are to study structure, beta decay properties, and
astrophysical processes.

Theoretical study for the first-forbidden $\beta $ decay in the lead region
assessed by the shell-model with the phenomenological interactions are
available in the literature
\cite{Warburton1990,Warburton1991,Warburton1991PRL,Suzuki2012,Zhi2013,Haselschwardt2020}.
The first-forbidden $\beta ^{-}$ decay of $^{206}$Tl$(0^{-})\to \,^{206}$Pb($0^{+}$)
in the framework of the shell-model was reported in Ref.~\cite{Warburton1990} by considering the core polarization effect. In this
work, they have found that the large quenching is needed in the couplings
constants. A more comprehensive study of the first-forbidden
$\beta $ decay of the mass number $A=205-212$ around the doubly magic nucleus
$^{208}$Pb was performed by Warburton \cite{Warburton1991} and extracted
the large mesonic enhancement factor
$\epsilon _{\mathrm{mec}}=2.01\pm 0.05$ in the rank-zero matrix element of
$\gamma _{5}$. In Ref.~\cite{Suzuki2012} the Gamow-Teller and first-forbidden
transitions were taken into account to calculate the half-lives of the
isotones with the neutron magic number of $N=126$. It was concluded that
the first-forbidden transitions are important to reduce the half-lives
in the study of the $r$-process including those besides the supernova explosions.
In Ref.~\cite{Zhi2013}, the study of the first-forbidden contributions
for the $r$-process waiting-point nuclei at the magic neutron numbers
$N=50$, 82, and 126 were reported using the large scale shell-model
calculations and also found that the large quenching factor is needed in
the coupling constants in the lead region. The high-precision theoretical
$\beta $ spectrum for the ground-state-to-ground-state first-forbidden
transitions of $^{212,214}$Pb that are relevant to the background of liquid
xenon dark matter detectors are reported in Ref.~\cite{Haselschwardt2020}. The structure of $^{208}$Po populated through
EC/$\beta ^{+}$ decay of $^{208}$At is reported in Ref.~\cite{Brunet}.

Recently, an experiment has been performed to study the $\beta ^{-}$ decay
of $^{207}$Hg into the $^{207}$Tl through $\gamma $-ray spectroscopy at
ISOLDE-CERN \cite{Berry2020}. In this experiment, they have observed several
new excited states and $\beta $ feeding branching ratio between the energy
range 2.6-4.0 MeV for $^{207}$Tl through $\beta ^{-}$ decay of
$^{207}$Hg. In this work \cite{Berry2020} shell-model results for the energy
spectra using KHM3Y interaction have been reported, although, no shell-model
results for the $\log ft$ values are reported. These new data motivated
us to perform large scale shell-model calculations for the $\log ft$ values
corresponding to $\beta ^{-}$ decay of $^{207}$Hg($9/2^{+}_{\mathrm{g.s.}}$)$
\to \,^{207}$Tl($J_{f}^{\pi }$) transitions. To the best of our knowledge,
there is no theoretical calculations are available for these transitions
in the literature. In the present study, we have made $1p$-$1h$ excitations
simultaneously for both protons and neutrons in the final states across
the doubly-shell closure $^{208}$Pb suggested by Warburton in Refs.~\cite{Warburton1990,Warburton1991,Warburton1991PRL}. The first-order contributions
from nucleon excitations across the core can significantly affect the first-forbidden
nuclear matrix elements \cite{Warburton1990}. In this work, we have adopted
the large quenching in the weak coupling constants, also the mesonic enhancement
factor $\epsilon _{\mathrm{mec}}=2.01\pm 0.05$
\cite{Warburton1991,Warburton1991PRL} in the rank-0 axial-charge matrix
element $\gamma _{5}$ corresponding to $\Delta J = 0$ transitions. The
present study will add more information to Berry et al.
\cite{Berry2020} work.

This paper is organized as follows. In Sec.~\ref{beta}, we give a short
overview of the formalism for the forbidden $\beta ^{-}$ decay. Results
and discussions about the interactions and $\beta $ decay rate calculations
are presented in the Sec.~\ref{results}. Finally, in Sec.~\ref{conclusions} we give summary and conclusions.

\section{$\beta $-decay theory}
\label{beta}

The partial half-life of the $\beta $ decay process can be expressed as
\cite{mika2017,behrens1982}
%
\begin{eqnarray}
\label{hf1}
t_{1/2}=\frac{\kappa }{\tilde{C}},
\end{eqnarray}
where $\kappa $ is the constant and has the value \cite{Patrignani}
\begin{eqnarray}
\kappa =
\frac{2\pi ^{3}\hbar ^{7}{\mathrm{ln(2)}}}{m_{e}^{5}c^{4}(G_{\mathrm{F}}{\mathrm{cos}}\theta _{\mathrm{C}})^{2}}=6289~s,
\end{eqnarray}
where the $\theta _{\mathrm{C}}$ is the Cabibbo angle. The $\tilde{C}$ is the
dimensionless integrated shape function which is defined as
%
\begin{eqnarray}
\label{tc}
\tilde{C}=\int _{1}^{w_{0}}C(w_{e})pw_{e}(w_{0}-w_{e})^{2}F_{0}(Z,w_{e})dw_{e}.
\end{eqnarray}
Where the $w_{0}$ is the end-point energy, $w_{e}$ and $p_{e}$ are the
total energy and momentum of the emitted electrons, respectively, and the
factor $F_{0}(Z,w_{e})$ is the Fermi function. The $C(w_{e})$ is the shape
factor, which contains all the nuclear structure information about the
$\beta $ decay transitions. To calculated the $ft$-values, we have multiplied
the partial half-life with the following dimensionless integrated Fermi
function
%
\begin{eqnarray}
f_{0}=\int _{1}^{w_{0}}pw_{e}(w_{0}-w_{e})^{2}F_{0}(Z,w_{e})dw_{e},
\end{eqnarray}
and the $\log ft$ value is calculated by
%
\begin{eqnarray}
\log ft=\log _{10}(f_{0}t_{1/2}[s]).
\end{eqnarray}
%
The general form of the shape factor $C(w_{e})$ in Eq.~\ref{tc} can be
expressed as
%
\begin{eqnarray}
\label{eq2}
C(w_{e}) &=& \sum _{k_{e},k_{\nu },K}\lambda _{k_{e}} \Big [M_{K}(k_{e},k_{\nu })^{2}+m_{K}(k_{e},k_{\nu })^{2}
-\frac{2\gamma _{k_{e}}}{k_{e}w_{e}}M_{K}(k_{e},k_{\nu })m_{K}(k_{e},k_{\nu })\Big ],
\end{eqnarray}
where the $K$ is the order of forbiddenness for $\beta $ decay, the indices
$k_{e}$ and $k_{\nu }$ ($k_{e}$, $k_{\nu }$ = 1,2,3,...) are the positive
integers that emerged from the partial wave expansion of the leptonic wave
functions. Here the auxiliary quantities are
$\gamma _{k_{e}}=\sqrt{k_{e}^{2}-(\alpha {Z})^{2}}$ and
$\lambda _{k_{e}}={F_{k_{e}-1}(Z,w_{e})}/{F_{0}(Z,w_{e})}$, where the
$\lambda _{k_{e}}$ is the Coulomb function and
${F_{k_{e}-1}(Z,w_{e})}$ is the generalized Fermi function
\cite{mst2006,mika2017,Anil2020PRC}. The quantities
$M_{K}(k_{e},k_{\nu })$ and $m_{K}(k_{e},k_{\nu })$ are the complicated expressions
of different nuclear matrix elements (NMEs) and other kinematical factors.

The NMEs contains all the nuclear-structure information in the form
%
\begin{eqnarray}[ll]
^{V/A}\mathcal{M}_{KLS}^{(N)}(pn)(k_{e},m,n,\rho )\nonumber\\
\quad {}=\frac{\sqrt{4\pi }}{\widehat{J}_{i}} \sum _{pn} \, ^{V/A}m_{KLS}^{(N)}(pn)(&k_{e},m,n,
\rho )(\Psi _{f}|| [c_{p}^{\dagger } \tilde{c}_{n}]_{K} || \Psi _{i}).
\label{eq:ME}
\end{eqnarray}

The nuclear matrix elements are divided into two parts: one part\\
$^{V/A}m_{KLS}^{(N)}(pn)(k_{e},m,n,\rho )$ is the single-particle matrix
elements (SPMEs), which characterizes the properties of the transition
operator and it is universal for all nuclear model. The SPMEs are calculated
in the present work from the harmonic-oscillator wave functions
\cite{mika2017,mst2006}. The second part
$(\Psi _{f}|| [c_{p}^{\dagger }\tilde{c}_{n}]_{K} || \Psi _{i})$ is the
one-body transition densities (OBTD) between initial $(\Psi _{i})$ and
final $(\Psi _{f})$ nuclear states. The OBTDs contain the nuclear-structure
information and they must be evaluated separately for each nuclear model.
The summation runs over the proton $(p)$ and neutron $(n)$ single-particle
states and the ``hat-notation'' reads
${\widehat{J}_{i}}=\sqrt{2J_{i}+1}$.

\section{Results and discussions}
\label{results}

\begin{figure}
\begin{center}
\includegraphics[width=9cm,height=7cm]{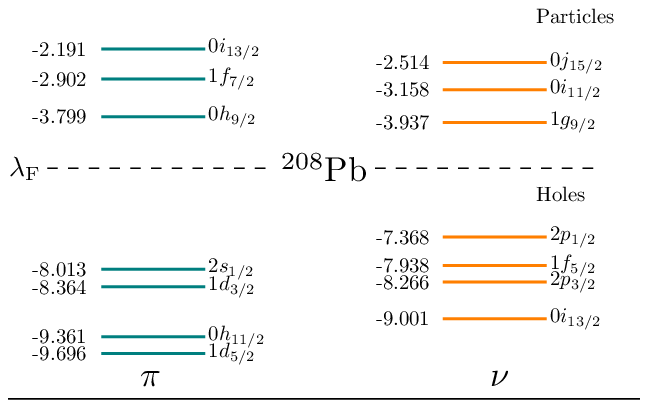}
\caption{The model space with the experimental single-particle energies that are
taken from the experiment energy spectra of $A=207$ and 209 relative to
$^{208}$Pb.}\label{md_spc}
\end{center}
\end{figure}

To calculate the one-body transition densities which are needed in the
nuclear matrix elements, we have performed theoretical calculations in
the framework of nuclear shell-model with KHH7B interaction. The KHH7B
interaction consists of the model space covering the shell gap in the ranges
$Z=58-114$ and $N=100-164$ around the $^{208}$Pb. The model space includes
the proton orbitals $1d_{5/2}$, $0h_{11/2}$, $1d_{3/2}$, and
$2s_{1/2}$ below $Z=82$ and the $0h_{9/2}$, $1f_{7/2}$, and
$0i_{13/2}$ above, and the neutron orbitals $0i_{13/2}$, $2p_{3/2}$,
$1f_{5/2}$, and $2p_{1/2}$ below $N=126$ and the $1g_{9/2}$,
$0i_{11/2}$, and $0j_{15/2}$ above. The orbitals used in this model space
are shown in Fig.~\ref{md_spc} with the experimental single particle/hole
energies relative to $^{208}$Pb. In KHH7B interaction, the cross shell
two-body matrix elements (TBMEs) were generated by the G-matrix potential
(H7B) \cite{Hosaka1985}, and the proton-neutron hole-hole and particle-particle
TBMEs are used from Kuo-Herling interaction \cite{KuoH1971} as modified
in \cite{War_Brown1991}. Previously, shell model results using KHH7B interaction
are reported in Refs.~\cite{Podolyak2015,Wilson2015,Berry2019,Wahid2020}. For the calculation,
we have used the nuclear shell-model code OXBASH \cite{Brown2004}.

\begin{figure}
\begin{center}
\hspace{-1cm}
\includegraphics[width=12cm,height=18.5cm]{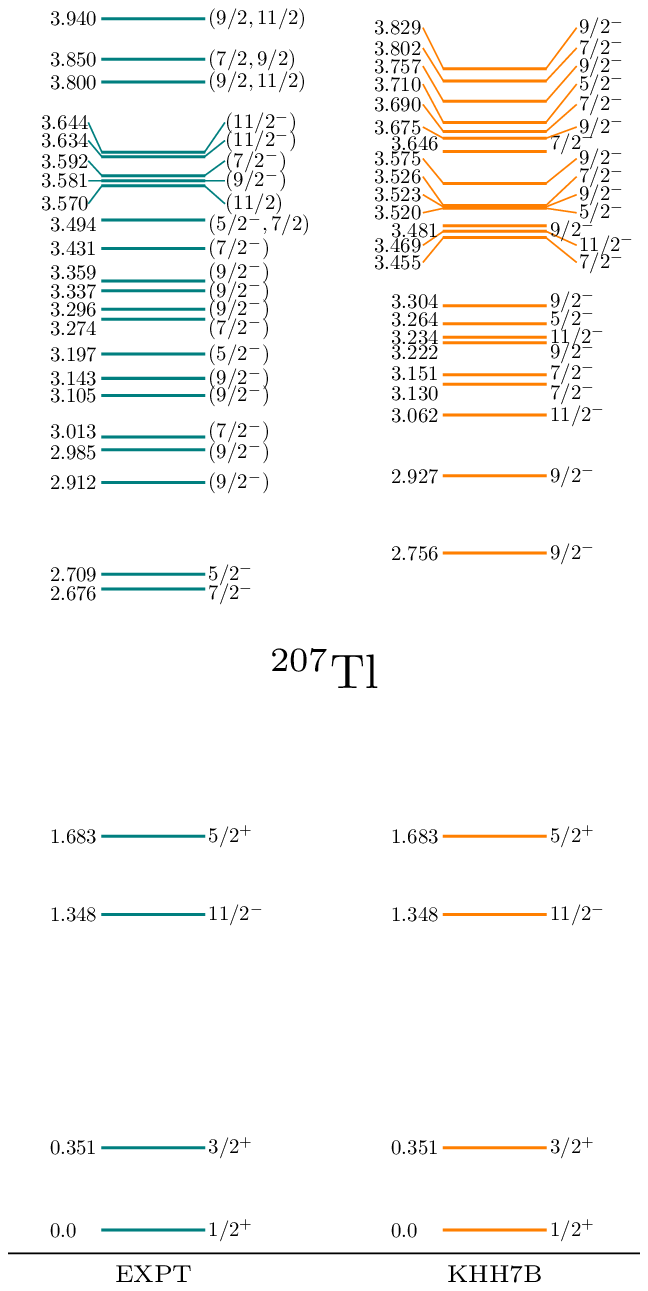}
\caption{The shell-model results for the low-lying energy spectra  for $^{207}$Tl with KHH7B  interaction in comparison with  the experimental data. }\label{fig:207Tl}
\end{center}
\end{figure}

In Fig.~\ref{fig:207Tl}, we have calculated the energy spectra of the
$^{207}$Tl and compared them with the experimental data. In the present
calculations, we have correctly reproduced ground state $1/2^{+}$ in
$^{207}$Tl from the KHH7B interaction. The ground state $1/2^{+}$ and few
excited states $3/2^{+}$, $11/2^{-}$, and $5/2^{+}$ are the single-particle
states that we have calculated without excitations of particle. In Table~\ref{table:conf}, we have shown the dominant wave functions of the various
low-lying energy states corresponding to $t=0$ (no excitation) and
$t=1$ (for both protons and neutrons excitation) calculations. The single
particle states $1/2^{+}$, $3/2^{+}$, $11/2^{-}$ and $5/2^{+}$ have reproduced
corresponding to the single-proton-hole states $\pi {s_{1/2}^{-1}}$,
$\pi {d_{3/2}^{-1}}$, $\pi {h_{11/2}^{-1}}$, and $\pi {d_{5/2}^{-1}}$, respectively,
as we can see in the Table~\ref{table:conf}. In the $\beta $ decay, a neutron
in $N<126$ orbitals of parent nucleus $^{207}$Hg decaying into the
$Z>82$ orbitals in the daughter nucleus $^{207}$Tl. In this situations,
both the proton and neutron pairs are broken across the core
$^{208}$Pb to form the $2p$-$3h$ configurations in the final states, by
making calculation with $1p$-$1h$ truncation for both protons and neutrons
across the $^{208}$Pb in the final state. Earlier, with this truncation,
the high-spin states in $^{207}$Tl are well-reproduced in Ref.~\cite{Wilson2015}. The nucleon excitation across the core has large influence
on the first-forbidden $\beta $ decay nuclear matrix elements
\cite{Warburton1990}. Mixing between the $t=0$ and core excited configuration
are blocked in our calculation for the energy spectra as shown in the Fig.~\ref{fig:207Tl}.

\begin{table}
\vspace{-1ex}

\caption{Dominant configurations of the states in $^{207}$Tl with KHH7B interaction.}
\label{table:conf}
\hspace{1cm}
\begin{tabular}{lclc}
\hline\hline
$J^{\pi}$ & $E$(MeV) & Dominant configuration & Probability\%\T\B \\\hline
$1/2^+$ & 0.000 & $\pi{s}_{1/2}^{-1}$ & 100\B\\
$3/2^+$ & 0.351 &  $\pi{d}_{3/2}^{-1}$ & 100\B\\
$11/2_1^-$ & 1.348 &  $\pi{h}_{11/2}^{-1}$ & 100\B\\
$5/2^+$ & 1.683 &  $\pi{d}_{5/2}^{-1}$& 100\B\\
$9/2^-_1$& 2.756 & $\pi{s}_{1/2}^{-2}\pi h_{9/2}$&61\B\\ 
$9/2^-_2$& 2.927 & $\pi{s_{1/2}^{-1}}\nu{p_{1/2}^{-1}}\nu{g_{9/2}}$&63\B\\
$11/2^-_2$  & 3.062 & $\pi{s_{1/2}^{-1}}\nu{p_{1/2}^{-1}}\nu{g_{9/2}}$&44 \B\\
$7/2^-_1$   & 3.130 &$\pi{s_{1/2}^{-1}}\nu{p_{3/2}^{-1}}\nu{g_{9/2}}$ &20\B\\
$7/2^-_2$   & 3.151 &$\pi{d_{3/2}^{-1}}\nu{p_{1/2}^{-1}}\nu{g_{9/2}}$& 35\B\\
$9/2^-_3$ &3.222& $\pi{s_{1/2}^{-1}}\nu{p_{1/2}\nu{g_{9/2}}}$&42\B\\
$11/2^-_3$  & 3.234  &$\pi{s_{1/2}^{-1}}\nu{p_{1/2}^{-1}}\nu{g_{9/2}}$&35\B \\
$5/2^-_1$  &3.264&$\pi{d_{3/2}^{-1}}\pi{s_{1/2}^{-1}}\pi{h_{9/2}}$&29\B \\
$9/2^-_4$ & 3.304&$\pi{s_{1/2}^{-1}}\nu{p_{1/2}^{-1}\nu{g_{9/2}}}$&35 \B\\
$7/2^-_3$   & 3.455 &$\pi{d_{3/2}^{-1}}\nu{p_{1/2}^{-1}}\nu{g_{9/2}}$& 37\B \\
$11/2^-_4$   & 3.469 &$\pi{s_{1/2}^{-1}}\nu{p_{1/2}^{-1}}\nu{i_{11/2}}$ & 51\B\\
$9/2^-_5$   & 3.481 &$\pi{d_{3/2}^{-1}}\nu{p_{1/2}^{-1}\nu{g_{9/2}}}$&36\B\\
$5/2^-_2$  &3.520&$\pi{d_{3/2}^{-1}}\nu{p_{1/2}^{-1}\nu{g_{9/2}}}$& 58\B\\
$9/2^-_6$   & 3.523 &$\pi{d_{3/2}^{-1}}\nu{p_{1/2}^{-1}\nu{g_{9/2}}}$&31\B\\
$7/2^-_4$ &3.526&$\pi{d_{3/2}^{-2}}\pi{h_{9/2}}$&26\B\\
$9/2^-_7$ &3.575&$\pi{s_{1/2}^{-1}}\nu{p_{3/2}^{-1}}\nu{g_{9/2}}$& 20\B\\
$7/2^-_5$&3.646&$\pi{s_{1/2}^{-1}}\nu{p_{3/2}^{-1}}\nu{g_{9/2}}$&31\B\\
$9/2^-_8$ &3.675&$\pi{s_{1/2}^{-1}}\nu{p_{1/2}^{-1}}\nu{i_{11/2}}$&32\B\\
$7/2^-_6$ & 3.690&$\pi{d_{3/2}^{-1}}\nu{p_{1/2}^{-1}}\nu{g_{9/2}}$&24 \B\\
$5/2^-_3$ &3.710&$\pi{d_{3/2}^{-2}}\pi{h_{9/2}}$&27\B\\
$9/2^-_9$ &3.757&$\pi{d_{3/2}^{-1}}\pi{s_{1/2}^{-1}}\pi{h_{9/2}}$& 20\B\\
$7/2^-_7$ &3.802&$\pi{s_{1/2}^{-1}}\nu{f_{5/2}^{-1}}\nu{g_{9/2}}$&21\B\\
$9/2^-_{10}$ &3.829&$\pi{s_{1/2}^{-1}}\nu{f_{5/2}^{-1}}\nu{g_{9/2}}$&53\B\\
\hline\hline
\end{tabular}
\end{table}

\begin{table*}
	\vspace{-1ex}
	\label{table:logft}
	\caption{Comparison of the theoretical and experimental $\log ft$ values  for the first-forbidden $\beta^-$ transitions of $^{207}$Hg ($9/2^+$) into the different excited states in $^{207}$Tl. The experimental excitation energy, branching fractions, and $\log ft$ values are taken from \cite{Berry2020}.}
	\hspace{-2cm}
	\begin{tabular}{lccccccc}
		\hline\hline
		&  & & &&&\multicolumn{2}{c}{$\log ft$}\\
		\cline{7-8}
		$J^\pi_f$(Expt.)&$J^\pi_f$(SM) & $E_i$(MeV) (Expt.) & $E_i$(MeV) (SM)& $Q$(MeV) &$\beta^-$ branching & Expt.& SM   \T\B\\
		\hline
		$11/2^-$    &   $11/2_1^-$              & 1.348  & 1.348 & 3.202    & 11(7)   &7.2(4) & 8.19\T\B\\
		$7/2^-$     & $7/2_1^-$  & 2.676   & 3.130 & 1.874  &0.3(2)     &7.8(3) & 6.24 \T\B\\
		$5/2^-$      & $5/2_1^-$     & 2.709   & 3.264  & 1.841 &0.02(2)&$>$8.7 & 9.12\T\B\\
		$(9/2^-)$    & $9/2_1^-$   & 2.912      & 2.756  &  1.638&6(2)    &6.3(2) & 6.62 \T\B\\
		$(9/2^-)$    & $9/2_2^-$ & 2.985     & 2.927 &  1.565 &40(5)   &5.42(7) &4.86\T\B\\
		$(7/2^-)$   & $7/2_2^-$& 3.013   & 3.151  &1.537 &0.21(5) &7.7(2) & 6.23 \T\B\\
		
		$(9/2^-)$   &$9/2_3^-$& 3.105    & 3.222  & 1.445 &21(3)        &5.58(8) & 5.59 \T\B\\
		$(9/2^-)$    &$9/2_4^-$ & 3.143  & 3.304   & 1.407 &8(1)           &5.95(7) & 5.90\T\B\\
		$(5/2^-)$    &$5/2_2^-$& 3.197 &  3.520   & 1.353&0.001(1)     &$>$9.5 & 9.86 \T\B\\
		$(7/2^-)$     &$7/2_3^-$ &  3.274 & 3.455   & 1.276  &2.3(3)         &6.34(8) &6.66 \T\B\\
		$(9/2^-)$    &  $9/2_5^-$& 3.296  & 3.481   & 1.254 &3.2(4)         &6.17(8)  & 6.81\T\B\\
		$(9/2^-)$     &$9/2_6^-$ & 3.337 & 3.523    & 1.213&6.5(6)           &5.81(7)  & 6.00\T\B\\
		$(9/2^-)$     &$9/2_7^-$ & 3.359 &  3.575    &  1.191&3.8(4)      &6.01(7) &  6.89\T\B\\
		$(7/2^-)$     &$7/2_4^-$& 3.431 &  3.526      &  1.119 &0.70(8)      &6.65(8) & 8.20\T\B\\
		$(5/2^-,7/2)$ &$5/2_3^-$   & 3.494  &   3.710  &1.056  &0.0060(9)&8.63(9) & 10.68\T\B\\
		$( 5/2^-, 7/2)$ & $7/2_5^-$ & 3.494  & 3.646   &1.056 &0.0060(9)&8.63(9) & 7.86\T\B\\
		$(11/2)$  &$11/2_2^-$     & 3.570 &  3.062    & 0.980& 0.12(2)      &7.21(10) &5.80\T\B\\
		$(9/2^-)$     &$9/2_8^-$ & 3.581 & 3.675      & 0.969 &0.20(2)     &6.97(8) & 7.99\T\B\\
		$(7/2^-)$ & $7/2_6^-$   & 3.592 & 3.690     & 0.958 &0.14(2)     &7.11(9) & 6.92\T\B\\
		$(11/2^-)$ &$11/2_3^-$     &3.634 & 3.234    & 0.916   &0.70(8)    &6.34(8) & 6.04\T\B\\
		$(11/2^-)$  &$11/2_4^-$    & 3.644 & 3.469     & 0.906     &0.28(4) &6.72(9)& 6.77 \T\B\\
		$( 9/2,11/2)$  &$9/2_9^-$ & 3.800 &  3.757    &0.750  &0.041(5)     &7.27(9) & 7.94 \T\B\\
	    $( 9/2,11/2)$  &$11/2_5^-$ & 3.800 &  3.523    &0.750  &0.041(5)     &7.27(9) & 8.85 \T\B\\
	    $(7/2, 9/2)$    &$7/2_{7}^-$ & 3.850 &   3.802        &0.700 &0.022(6)      &7.4(2) & 6.38\T\B\\
		$(7/2, 9/2)$    &$9/2_{10}^-$ & 3.850 &  3.829   &0.700 &0.022(6)      &7.4(2) & 8.16\T\B\\
		$(9/2,11/2)$  & $9/2_{11}^-$& 3.940  &  3.876     &0.610  &0.031(4)      &7.08(10) & 7.86\T\B\\
		$(9/2,11/2)$  & $11/2_{6}^-$& 3.940  &  3.624     &0.610  &0.031(4)      &7.08(10) & 6.58\T\B\\
		
		\hline\hline
	\end{tabular}
\end{table*}

Once the energy spectra are reasonably reproduced, then we have used the
wave function to calculate the nuclear matrix elements needed in the
$\beta $ decay rate. We have considered the two main aspects in our calculations:
First is the effective values of weak coupling constants and the second
one is the effect of mesonic enhancement in the axial-charge matrix element
$\gamma _{5}$ for the $\Delta {J}=0$ transitions. First, we will discuss
the quenching in the coupling constants that are playing an important role
in the $\beta $ decay rates. In the previous study of the lead region,
heavily quenched values of the coupling constants $g_{\mathrm{A}}$ and
$g_{\mathrm{V}}$ for the first-forbidden $\beta $ decay of $1^{-}$ transitions
were suggested in Refs.~\cite{Warburton1991,Warburton1990,Suzuki2012,Zhi2013}. Warburton's calculation
for the first-forbidden transitions in the lead region has found that the
quenching factor in the coupling constants
$g_{\mathrm{A}}/{g_{\mathrm{A}}^{\mathrm{free}}}\approx 0.47$ and
$g_{\mathrm{V}}/{g_{\mathrm{V}}^{\mathrm{free}}}\approx 0.64$ \cite{Warburton1990}. Two
set of quenching factors, $g_{\mathrm{A}}/{g_{\mathrm{A}}^{\mathrm{free}}}=0.34$,
$g_{\mathrm{V}}/{g_{\mathrm{V}}^{\mathrm{free}}}=0.67$ and
$g_{\mathrm{A}}/{g_{\mathrm{A}}^{\mathrm{free}}}=0.51$,
$g_{\mathrm{V}}/{g_{\mathrm{V}}^{\mathrm{free}}}=0.30$ are calculated for the first-forbidden
$\beta ^{-}$ decay of
$^{205}{\mathrm{Tl}}(1/2^{+})\to \,^{205}{\mathrm{Pb}}(1/2^{-})$
\cite{Rydstrom1990}. Recently, Zhi \cite{Zhi2013} et al., have performed
the large-scale shell-model calculations for the number of nuclei around
the magic numbers $N=82,126$ for both Gamow-Teller and first-forbidden
$\beta $-decay and found that the quenching factor in coupling constants
are $g_{\mathrm{A}}/g_{\mathrm{A}}^{\mathrm{free}}=0.38$ and
$g_{\mathrm{V}}/{g_{\mathrm{V}}^{\mathrm{free}}}=0.51$. From this literature survey, it
is concluded that the value of vector and axial-vector coupling constants
are strongly quenched in the heavier mass region. This might be due to
heavily truncated shell-model calculations in the heavier mass region.

The effect of the mesonic enhancement of the axial-charge matrix element
$\gamma _{5}$ were studied in the lead region for several $0^{-}$ transitions
and suggested the value of enhancement factor
$\epsilon _{\mathrm{mec}}=2.01\pm 0.05$ that is about 100\% of the impulse approximation
\cite{Warburton1991}. For the further calculation of the $\log ft$ values,
we have used recently calculated quenching factor of
$g_{\mathrm{A}}/g_{\mathrm{A}}^{\mathrm{free}}$=0.38 and
$g_{\mathrm{V}}/g_{\mathrm{V}}^{\mathrm{free}}$=0.51 by Zhi et al. \cite{Zhi2013} and along
with this we have used the mesonic enhancement factor
$\epsilon _{\mathrm{mec}}=2.01\pm 0.05$ in the matrix element
$\gamma _{5}$ for the $0^{-}$ transition \cite{Warburton1991}.

To calculate the $\beta $ decay rate, a precise $Q$ value is needed. In
our calculations, we have adopted the experimental $Q=4.550$ MeV
\cite{Wang2017} value to calculate the $\log ft$ value for the
$\beta ^{-}$ decay of $^{207}$Hg. Our theoretical prediction for the
$\log ft$ values of $^{207}$Hg are compared with the recent experimental
data \cite{Berry2020} in Table~\ref{table:logft} together with the experimental
excitation energy of final states and $Q$ values. The theoretical prediction
of $\log ft$ values for the studied transition are not available in the
literature. First time we have calculated the $\log ft$ from the nuclear
shell-model by including the core polarization effect in the nuclear matrix
elements. Our calculated $\log ft$ values are found in good agreement with
the experimental data.

The state $11/2_{1}^{-}$ at 1.348 MeV has the single particle character
that is reproduced by the one-proton-hole ($\pi {\mathrm{h}}_{11/2}^{-1}$) configuration
with 100\% occupancy in $t=0$ truncation. If we consider the $2p-3h$ admixture
in the $11/2_{1}^{-}$ for the $\beta $ decay study then the
$11/2_{1}^{-}$ is predicted with probability 83\%
${\pi \mathrm{h}}_{11/2}^{-1}$. In Table~\ref{table:logft}, we have reported
the $\log ft$ value of 8.19 corresponding to the $11/2^{-}$ state with
the configuration mixing in the final state, this is close to the experimental
value of 7.2(4). On the other hand without excitation i.e. for $t=0$ calculation,
we have obtained the $\log ft$ value 8.52.

Our theoretical prediction of $\log ft$ values for the $5/2^{-}$ states
at 2.709 MeV and 3.197 MeV are in reasonable agreement with the experimental
data. These states could be a $5/2^{-}$ and obtained $\log ft$ values are
9.12 and 9.86 corresponding to energy value 2.709 MeV and 3.197 MeV, respectively,
this is consistent with the unique first-forbidden decay. A state at 2.912
MeV is experimentally tentative ($9/2^{-}$), with $\log ft$ value 6.3(2).
Our calculated $\log ft$ for $9/2_{1}^{-}$ is 6.62, which is in excellent
agreement with the experimental data. The calculated $\log ft$ values for
$9/2_{3}^{-}$ and $9/2_{4}^{-}$ are 5.59 and 5.90, respectively, corresponding
to the experimental values 5.58(8) and 5.95(7). For other $9/2^{-}$ states
at 3.296, 3.337 and 3.359 MeV the calculated $\log ft$ values are in good
agreement with the experimental data.

In Refs.~\cite{Berry2020,Berry2019}, they have experimentally populated
new states $(5/2^{-},7/2)$ at 3.494 MeV with the $\log ft$ value of 8.63(9),
and this state is identical to $7/2^{+}$ at 3.474 MeV observed previously
in \cite{Kondev2011}. In Ref.~\cite{Berry2019} the experimentally calculated
$\log ft$ value for the 3.474 state is 8.53(9) and this is exactly predicted
by the shell-model calculation with the $2p-3h$ admixture. Transitions
from $9/2^{+}$ of $^{207}$Hg to ($5/2^{-}$ and $7/2^{-}$) of
$^{207}$Tl (3.494 MeV) are characterized as the first-forbidden unique
and nonunique $\beta ^{-}$ decay, respectively. Thus we have performed
two sets of calculations for $9/2^{+}\to \,7/2^{-}$ and
$9/2^{+}\to \,5/2^{-}$ transitions, the calculated $\log ft$ values corresponding
to the first-forbidden unique and nonunique $\beta ^{-}$ decay are 10.68
and 7.86, respectively. Here, the $\log ft$ values with the transition
$9/2^{+}\to \,7/2^{-}$(3.494  MeV) is more close to the experimental
result. Also, our theoretical prediction of the $\log ft $ value for the
transition $9/2^{+}\to \,5/2^{-}$ ~(3.494  MeV) is found to be consistent
with the first-forbidden unique $\beta $ decay.

The 3.570 MeV state is experimentally identified as the (11/2) with
$\log ft$ value 7.21(10). The calculated $\log ft$ value for
$11/2^{-}$ state is 5.80. It can be seen that this transition is predicted
too fast as compared with the experimental data. The spin and parity of
states at the 3.800, 3.850, and 3.940 MeV are not confirmed from the experimental
side. In Ref.~\cite{Berry2020}, they have given some predictions of spin
only at these energies with $\log ft$ values. Hence, we have calculated
the $\log ft$ values for these three states from the nuclear shell-model
corresponding to $9/2^{-}$ and $11/2^{-}$ states at 3.800 MeV,
$7/2^{-}$ and $9/2^{-}$ states at 3.850 MeV and $9/2^{-}$ and
$11/2^{-}$ states at 3.940 MeV. So our theoretical prediction could be
useful for future experiments to confirm the spin and parity of these three
states at 3.800, 3.850, and 3.940 MeV. In general, the computed
$\log ft$ values for all the studied transitions are in good agreement
with the experimental data. The calculated $11/2^{-}$ states corresponding
to experimental state at 3.570, 3.634, 3.644, 3.800 and 3.940 MeV are at
3.062, 3.234, 3.469, 3.523 and 3.624 MeV, respectively.

From the comparison of the experimental and the shell model results, the
octupole states are not well reproduced. Because the model space of KHH7B
is too small for collective octupole states. This explains the lower predicted
$log ft$ values for the lowest energy $5/2^{-}$ and $7/2^{-}$ states. Maybe
if we use KHM3Y interaction developed for a model space containing 24 orbitals
\cite{Brown2000}, then it will possible to reproduce octupole states in
our calculation.

\section{Conclusions}
\label{conclusions}

In this work, we have performed the shell-model calculations for the
$\log ft$ values using KHH7B interaction for the first-forbidden
$\beta ^{-}$ transitions corresponding to recently available experimental
data \cite{Berry2020}. In these calculations, we have used the $2p$-$3h$
admixture in the final state to include the core polarization effects in
the $\beta $ decay study. First, to test the predictive power of the used
Hamiltonian, we have calculated the energy spectrum. After that, we have
used the wave function in the calculation of the nuclear matrix elements
that are needed in the $\beta $ decay rate calculations. Also, we consider
the mesonic enhancement factor $\epsilon _{\mathrm{mec}}=2.01\pm 05$ in the
axial-charge matrix element $\gamma _{5}$ and the quenching factor in coupling
constants suggested in the literature in the lead region. First time we
have calculated the $\log ft$ values from the shell-model with well-known
interaction and found a reasonable agreement with the recent experimental
data. We have also given the prediction of some unknown states based on
the calculated $\log ft$ values. Our predictions might be very useful as
input for future experiment. It is also noticed that calculated energy
spectra for high-spin states are compressed thus further theoretical calculations
with more particle excitations are needed to reproduce the collectivity
and achieve convergence in the energy spectra for these states.%

\section*{ACKNOWLEDGMENTS}
A. K. would like to thank  the Ministry of Human Resource Development (MHRD), Government of India, for the financial support for his thesis work.  P.C.S. acknowledges a research grant from SERB (India),
CRG/2019/000556. We would like to thank Prof. Zs. Podoly{\'{a}}k  for several useful discussions in this work.

%

\end{document}